%Paper: hep-th/9211115
%From: P.E.Dorey@newton.cam.ac.uk
%Date: Tue, 24 Nov 92 21:03 GMT

%
%%%%%%%%%%%%%%%%%%%%%%%%%%%%%%%%%%%%%%%%%%%%%%%%%%%%%%%%%%%%%
%
%  PlainTeX format, uses the harvmac macro package.
%
%  Two postscript figures accompany this paper, called
%  staircase_fig1 and staircase_fig2. They have been
%  submitted through the new 'fig' command, and are
%  intended to be printed separately from the main
%  text of the paper.
%
%%%%%%%%%%%%%%%%%%%%%%%%%%%%%%%%%%%%%%%%%%%%%%%%%%%%%%%%%%%%%
%
\input harvmac
%
%%%%%%%%%%%%%%%%%%%%%%%%%%%%%%%%%%%%%%%%%%%%%%%%%%%%%%%%%%%%%
%
%  Modification to Harvmac's reference listing
%
%\def\listrefs{\footatend\vfill\supereject\immediate\closeout\rfile\writestoppt
%\baselineskip=12pt\parskip=2pt plus 1pt
%\noindent{{\bf References}}\medskip{\frenchspacing%
%\parindent=20pt\escapechar=` \input refs.tmp\vfill\eject}\nonfrenchspacing}
%
%%%%%%%%%%%%%%%%%%%%%%%%%%%%%%%%%%%%%%%%%%%%%%%%%%%%%%%%%%%%%%
%
%  Modification to Harvmac's title command
%
\def\Title#1#2{\nopagenumbers\abstractfont\hsize=\hstitle\rightline{#1}%
\vskip .6in\centerline{\titlefont #2}\abstractfont\vskip .5in\pageno=0}
%
%%%%%%%%%%%%%%%%%%%%%%%%%%%%%%%%%%%%%%%%%%%%%%%%%%%%%%%%%%%%%%
%
%  These starting defns add a couple of things to harvmac,
%  so that referencing needn't worry about the first ref/
%  subsequent ref distinction. Latest harvmacs get round this
%  by some other route, but life is too short to go changing
%  everything now...
%
\def\RF#1#2{\if*#1\ref#1{#2.}\else#1\fi}
\def\NRF#1#2{\if*#1\nref#1{#2.}\fi}
\def\refdef#1#2#3{\def#1{*}\def#2{#3}}
%
%%%%%%%%%%%%%%%%%%%%%%%%%%%%%%%%%%%%%%%%%%%%%%%%%%%%%%%%%%%%%%
%
\def\ts{\hskip .16667em\relax}

\def\CMP{{\it Comm.\ts Math.\ts Phys.\ts}}

\def\IJMP{{\it Int.\ts J.\ts Mod.\ts Phys.\ts}}
\def\JMP{{\it J.\ts Math.\ts Phys.\ts}}
\def\JP{{\it J.\ts Phys.\ts}}
\def\MPL{{\it Mod.\ts Phys.\ts Lett.\ts}}
\def\NP{{\it Nucl.\ts Phys.\ts}}
\def\PL{{\it Phys.\ts Lett.\ts}}

\def\PRL{{\it Phys.\ts Rev.\ts Lett.\ts}}

\def\SJNP{{\it Sov.\ts J.\ts Nucl.\ts Phys.\ts}}

\def\Zm{Zamolodchikov}
\def\AZm{A.\ts B.\ts \Zm}
\def\AlZm{Al.\ts B.\ts \Zm}
\def\me{P.\ts E.\ts Dorey}
\def\dur{H.\ts W.\ts Braden, E.\ts Corrigan, \me\ and R.\ts Sasaki}
%
%%%%%%%%%%%%%%%%%%%%%%%%%%%%%%%%%%%%%%%%%%%%%%%%%%%%%%%%%%%%%%
%
\refdef\rABLa\ABLa{C.\ts Ahn, D.\ts Bernard and A.\ts LeClair, \NP
{\bf B346} (1990) 409}

\refdef\rBCa\BCa{R. Brunskill and A. Clifton-Taylor, {\it English Brickwork}
 (Hyperion 1977)}

\refdef\rBCDSc\BCDSc{\dur,
``Affine Toda field theory and exact S-matrices",
\NP {\bf B338} (1990) 689}

\refdef\rBLa\BLa{D.\ts Bernard and A.\ts LeClair, \lq\lq Quantum group
 symmetries and non-local currents in 2D QFT",
 \CMP {\bf 142} (1991) 99}

\refdef\rBLb\BLb{D.\ts Bernard and A.\ts LeClair, \lq\lq The
fractional supersymmetric sine-Gordon models",
\PL {\bf B247} (1990) 309}

\refdef\rBNYa\BNYa{J.\ts Bagger, D.\ts Nemeschansky and S.\ts
Yankielowicz,
``Virasoro algebras with central charge $c>1$",
\PRL {\bf 60} (1988) 389}

\refdef\rBRa\BRa{V.\ts V.\ts Bazhanov and N.\ts Reshetikhin, \JP {\bf
A23} (1990) 1477}

\refdef\rCRa\CRa{P.\ts Christe and F.\ts Ravanini,
``$G_N\otimes G_L/G_{N+L}$ conformal field theories and their modular
invariant partition functions",
\IJMP {\bf A4} (1989) 897}

\refdef\rCSSa\CSSa{C.\ts Crnkovi\'c, G.\ts M.\ts Sotkov and M.\ts
Stanishkov,
``Renormalization group flow for general $SU(2)$ coset models",
\PL {\bf B226} (1989) 297}

\refdef\rDf\Df{\me, \NP {\bf B358} (1991) 654; \NP {\bf B374} (1992) 741}

\refdef\rDRa\DRa{\me\ and F.\ts Ravanini,
``Staircase models from affine Toda field theory",
preprint SPhT/92-065, DFUB-92-09, \IJMP {\bf A} in press}

\refdef\rFb\Fb{M.\ts D.\ts Freeman,
``On the mass spectrum of affine Toda field theory",
\PL {\bf B261} (1991) 57}

\refdef\rFZa\FZa{V.\ts A.\ts Fateev and \AZm, {\it Int. J. Mod. Phys.}
 {\bf A5} (1990) 1025}

\refdef\rFZb\FZb{V.\ts A.\ts Fateev and \AlZm,
 `Integrable perturbations of $Z_N$ parafermion models and the $O(3)$
 sigma model'
 \PL {\bf B271} (1991) 91}

\refdef\rFZc\FZc{V.\ts A.\ts Fateev and \AZm, \NP {\bf B280} 644}

\refdef\rGKOa\GKOa{P. Goddard, A. Kent and D. Olive, \PL {\bf B152} (1985) 88;
 \CMP {\bf 103} (1986) 105}

\refdef\rKg\Kg{A.\ts N.\ts Kirillov, {\it Zapiski Nauch. Semin. LOMI}
{\bf 164} (1987) 121}

\refdef\rKIKKMOa\KIKKMOa{Y.\ts Kitazawa, N.\ts Ishibashi, A.\ts Kato,
K.\ts Kobayashi, Y.\ts Matsuo and S.\ts Odake,
``Operators product expansion coefficients in $N=1$ superconformal
theory and slightly relevant perturbation",
\NP {\bf B306} (1988) 425}

\refdef\rKMa\KMa{T.\ts R.\ts Klassen and E.\ts Melzer, ``Purely
elastic scattering theories and their ultraviolet limits", \NP {\bf B338}
(1990) 485}

\refdef\rKMb\KMb{T.\ts R.\ts Klassen and E.\ts Melzer, ``The
thermodynamics of purely elastic scattering theories and conformal
perturbation theory", \NP {\bf B350} (1991) 635}

\refdef\rKMf\KMf{T.\ts Klassen and E.\ts Melzer, ``Spectral flow
between conformal field theories in $1{+}1$ dimensions",
\NP {bf B370} (1992) 511}

\refdef\rKMQa\KMQa{D.\ts Kastor, E.\ts Martinec and Z.\ts Qiu,
``Current algebra and conformal discrete series",
\PL {\bf B200} (1988) 434}

\refdef\rKMSa\KMSa{D.\ts Kastor, E.\ts Martinec and S.\ts Shenker,
``RG flow in $N=1$ discrete series",
\NP {\bf B316} (1989) 590}

\refdef\rLc\Lc{L.\ts Lewin, Dilogarithms and associated functions
(Macdonald 1958)}

\refdef\rLd\Ld{M.\ts L\"assig,
``Multiple crossover phenomena and scale hopping in two dimensions",
\NP {\bf B380} (1992) 601}

\refdef\rLe\Le{M.\ts L\"assig,
``Exact critical exponents of the staircase model",
J\"ulich preprint (February 1992)}

\refdef\rLCa\LCa{A.\ts W.\ts W.\ts Ludwig and J.\ts L.\ts Cardy,
``Perturbative evaluation of the conformal anomaly at new critical
points with applications to random systems",
\NP {\bf B285} (1987) 687}

\refdef\rMk\Mk{M.\ts J.\ts Martins,
``The Thermodynamic Bethe Ansatz for deformed $WA_{N-1}$ conformal
field theories",
\PL {\bf B277} (1992) 301}

\refdef\rMl\Ml{M.\ts J.\ts Martins,
``Renormalization group trajectories from resonance factorized
S-matrices",
preprint SISSA-EP-72, \PRL in press\semi
M.\ts J.\ts Martins,
``Exact resonance A-D-E S-matrices and their renormalization group
trajectories",
preprint SISSA-EP-85}

\refdef\rMo\Mo{M.\ts J.\ts Martins,
``Analysis of asymptotic conditions in resonance functional
hierarchies",
preprint SISSA-EP-151}

\refdef\rMp\Mp{M.\ts J.\ts Martins,
``RG flows and resonance scattering amplitudes",
preprint SISSA-EP-154}

\refdef\rPg\Pg{R.\ts G.\ts Pogosyan,
``Study of the neighbourhoods of superconformal fixed points in
two-dimensional field theory",
\SJNP {\bf 48} (1988) 763}

\refdef\rRb\Rb{F.\ts Ravanini, ``Thermodynamic Bethe Ansatz for ${\cal
G}_k\otimes{\cal G}_l/{\cal G}_{k+l}$ coset models perturbed by their
$\phi_{1,1,Adj}$ operator",
\PL {\bf B282} (1992) 73}

\refdef\rRc\Rc{F.\ts Ravanini,
``An infinite class of new conformal field theories with extended
algebras",
\MPL {\bf A3} (1988) 397}

\refdef\rRTVa\RTVa{F.\ts Ravanini, R.\ts Tateo and A.\ts Valleriani,
``Dynkin TBAs",
preprint DFUB-92-11, DFTT-31/92}

\refdef\rYYa\YYa{C.\ts N.\ts Yang and C.\ts P.\ts Yang, \JMP {\bf 10}
(1969) 1115}

\refdef\rZb\Zb{\AZm, {\it Int. J. Mod. Phys.} {\bf A4} (1989) 4235}

\refdef\rZc\Zc{\AZm, {\it JETP Letters} {\bf 43} (1986) 730}

\refdef\rZd\Zd{\AZm, {\it Int. J. Mod. Phys.} {\bf A3} (1988) 743}

\refdef\rZe\Ze{\AZm, {\it Sov. Sci. Rev., Physics}, {\bf v.2} (1980)}

\refdef\rZf\Zf{\AZm, {\it Teor. Mat. Fiz.} {\bf 65} (1985) 347}

\refdef\rZg\Zg{\AlZm,
 ``Thermodynamic Bethe Ansatz in Relativistic Models. Scaling 3-state
 Potts and Lee-Yang Models",
 \NP {\bf B242} (1990) 695}

\refdef\rZk\Zk{\AlZm,
 ``Thermodynamic Bethe Ansatz for RSOS scattering theories",
 \NP {\bf B358} (1991) 497}

\refdef\rZl\Zl{\AlZm,
 ``From tricritical Ising to critical Ising by Thermodynamic Bethe
 Ansatz",
 \NP {\bf B358} (1991) 524}

\refdef\rZn\Zn{\AlZm,
 ``TBA equations for integrable perturbed\hfil\break
 $SU(2)_k\times $$SU(2)_l / SU(2)_{k+l}$ coset models",
 \NP {\bf B366} (1991) 122}

\refdef\rZo\Zo{\AlZm,
``Resonance factorized scattering and roaming trajectories", preprint
ENS-LPS-335, 1991}

\refdef\rZp\Zp{\AZm, \SJNP {\bf 44} (1986) 529}

\refdef\rZq\Zq{\AZm,
``Renormalization group and perturbation theory about fixed points in
two-dimensional field theory",
\SJNP {\bf 46} (1987) 1090}

\refdef\rZz\Zz{\Za\semi\Zb}
%
%%%%%%%% random definitions %%%%%%%%%%%%%%%%
\def\w{s}
\def\mod{\hbox{mod}}
\def\bar{\overline}
\def\hat{\widehat}
\def\tilde{\widetilde}
\def\({\left(}
\def\){\right)}
\def\[{\left[}
\def\]{\right]}

\def\th{^{\rm th}}

\def\exp{{\rm exp}}

\def\CG{{\cal G}}

\def\Z{{\bf Z}} \def\R{{\bf R}} 
\def\ZK{\Z$_k$}

\def\ep{\varepsilon}   \def\p{\phi}
\def\t{\theta}

\def\rdilog{{\cal L}}
\def\strutdepth{\dp\strutbox} % see Texbook ex 14.28
\def\probsymbol{\vtop to \strutdepth{\baselineskip\strutdepth
  \vss\llap{\bf ??~~}\null}}
\def\prob#1{\strut\vadjust{\kern-\strutdepth\probsymbol}{\bf #1}}
\def\coset#1#2{$\CG^{(#1)}{\times}\CG^{(#2)}/\CG^{(#1+#2)}$}
\def\ccoset#1#2#3{$\CG^{(#1)}{\times}\CG^{(#2)}/\CG^{(#3)}$}
\def\intt{\int_{-\infty}^{\infty}}
\def\forget#1{}
\def\AlZ{Al.\ts Zamolodchikov}
\def\newt{\centerline{Isaac Newton Institute for Mathematical
Sciences,}
\centerline{20 Clarkson Road, Cambridge CB3 0EH, UK}}
%
%%%%% diagram-drawing macros %%%%%%%%%%%%%%%%%%%%%%%%%%%%%%%%%%%%%%%
\def\clap#1{\hbox to 0pt{\hss #1\hss}}
\def\label#1{\raise3ex\clap{$\scriptstyle #1$}}
\def\massive#1{\hskip 4.5pt\clap{$\bigotimes$}\label{#1}\hskip 4.5pt}
\def\massless#1{\hskip 4.5pt\clap{$\bigcirc$}\label{#1}\hskip 4.5pt}
\def\link{------}
\def\dlink{-- -- -- -- --}
\def\sdlink{-- -- --}
\def\ssdlink{-- --}
%%%%%%%%%%%%%%%%%%%%%%%%%%%%%%%%%%%%%%%%%%%%%%%%%%%%%%%%%%%%%%%%%%%%%
%
% paper starts here
%
\Title{\vbox{\baselineskip12pt\hbox{CERN-TH.6739/92}%
\hbox{NI\ts 92009}\hbox{DFUB-92-21}\hbox{hep-th/9211115}}}%
{Generalising the staircase models}
\centerline{Patrick Dorey}
\smallskip
{\footnotefont
\newt\centerline{\it and}
\centerline{CERN TH, 1211 Geneva 23, Switzerland}
\centerline{\tt dorey@surya11.cern.ch}
}
\bigskip
\centerline{Francesco Ravanini}
\smallskip
{\footnotefont
\newt\centerline{\it and}
\centerline{INFN -- Sez. di Bologna,}
\centerline{Via Irnerio 46, I-40126 Bologna, Italy}
\centerline{\tt ravanini@bologna.infn.it}
}
\vskip .3in
\baselineskip=15pt plus 2pt minus 1pt
Systems of integral equations are proposed which generalise those
previously encountered in connection with the so-called staircase models.
Under the assumption that these equations describe the finite-size
effects of relativistic field theories via the Thermodynamic Bethe
Ansatz, analytical and numerical evidence is given for the existence
of a variety of new roaming renormalisation group trajectories. For each
positive integer $k$ and $\w=0,\dots, k{-}1$, there is a one-parameter
family of trajectories, passing close by the coset conformal field
theories \ccoset k {nk+\w} {(n+1)k+\w}\ before finally flowing to a
massive theory for $\w{=}0$, or to another coset model for $\w{\neq}0$.

\bigskip
% \draft
\Date{\vbox{\baselineskip12pt\hbox{CERN-TH.6739/92}%
\hbox{November 1992}}}

\newsec{Introduction}

The staircase model of \AlZ\ts\RF\rZo\Zo\ is a simple relativistic
factorized scattering
theory in 1+1 dimensions, which shows signs of a very non-trivial
renormalisation group  behaviour. It describes a single boson
with mass $m$ and two-particle scattering amplitude
\eqn\zamS{S(\t)=\tanh\({\t-\t_0\over 2}-{i\pi\over 4}\)
\tanh\({\t+\t_0\over 2}-{i\pi\over 4}\)\, ,}
where $\t_0$ is a real parameter. Assuming the existence of an
underlying field theory, the model can be studied at all
length-scales by placing it on a cylinder and
varying the circumference $R$. In particular, the ground-state
scaling function $c(x)$ ($x=\log{mR\over 2}$)
can be obtained by means of the Thermodynamic Bethe Ansatz
(TBA)\ts\NRF\rYYa\YYa\NRF\rZg\Zg\refs{\rYYa,\rZg}.
(The ground-state scaling function can be interpreted as an `effective
central charge' for the non scale-invariant theory, and is
related to the vacuum Casimir energy
by $E(R)=-\pi c(x)/6R\,$; for a unitary scale-invariant theory it
is just equal to the central charge.) Plotting $c(x)$ as
a function of $x$ shows a series of plateaux,
connected by steps each time $x$ is near an
integer or half-integer multiple of $\t_0$. This staircase-like
pattern becomes more pronounced as $\t_0$ increases,
the values taken by $c(x)$ on the plateaux
then running through the series $c_p=1-6/p(p{+}1)$, the central charges
of the $c{<}1$ minimal models.
This suggests an RG flow from an ultraviolet fixed point
($R\rightarrow 0$; $x\rightarrow -\infty$) with $c=1$, to a
trivial $c=0$ fixed point in the infrared,
passing close by each minimal model ${\cal M}_p$ in turn.
Varying $\t_0$ results in a one-parameter family of such
flows; the larger $\t_0$, the more closely the trajectory visits each
minimal model in its journey from $c=1$ to $c=0$.

{}From another point of view, A.\ts B.\ts Zamolodchikov\ts\RF\rZq\Zq,
and Ludwig and Cardy\ts\RF\rLCa\LCa\ showed some time ago that for
large $p$ the deformation of ${\cal M}_p$
by its $\phi_{13}$ operator leads in the infrared to ${\cal M}_{p-1}$,
so long as the coupling constant is positive. More recently, TBA analysis
by \AlZ\ts\NRF\rZl{\Zk\semi\Zl}\NRF\rZn\Zn\refs{\rZl,\rZn},
applicable for all values of $p$,
has reinforced this picture. The perturbed model, commonly
denoted ${\cal M}A_p^{(+)}$, can thus be associated with an RG flow that
hops between two neighbouring minimal models, and the
staircase model can be seen as an approximation to this whole series of
hopping flows, the approximation becoming better as $\t_0$
increases.

The minimal models are only the first of many infinite
series of conformal field theories that can be constructed by means of
the GKO coset construction\ts\RF\rGKOa\GKOa. For example, given any
simple simply-laced Lie algebra $\CG= A,$ $D$ or $E$, there is a
series of $W_{\CG}$-minimal models, rational conformal field theories
described by the coset \coset 1 l. Since the minimal models are
recovered by chosing $\CG=A_1$, it is natural to ask whether
staircase models can be found for the other $W_{\CG}$ series. The
fact that Zamolodchikov's S-matrix \zamS\ is an analytic continuation
in the coupling constant of the sinh-Gordon S-matrix
(already remarked in ref.\ts\rZo) provides a strong hint
that the required generalisation is to be found in the analytic
continuation of the real-coupling affine Toda S-matrices, and this
turns out to be the case\ts\NRF\rMl\Ml\NRF\rDRa\DRa\refs{\rMl,\rDRa}.
To complete the analogy with the minimal sequence, it
can be argued that `one-hop' flows similar to those associated with
the models ${\cal M}A^{(+)}_p$ should also exist for each
$W_{\CG}$ series. The generalisations of
Zamoldochikov's $A_1$ TBA analysis\ts\rZl\ to arbitrary
$\CG$\ts\NRF\rMk\Mk\NRF\rRb\Rb\refs{\rMk,\rRb}\ lend support to this view.

However this is by no means the end of the story.
In particular, there are the models
\coset k l\ts\RF\rCRa\CRa. For each fixed $k$, they form a series with
central charges $c_l=r(h{+}1)kl(k{+}2h{+}l)/(h{+}k)(h{+}l)(h{+}k{+}l)$,
where $r$ is the rank of $\CG$ and $h$ the Coxeter number.
In all cases, the operator analogous to $\phi_{13}$ is
$\phi_{id,id,adj}$, the three indices labelling particular
representations of $\CG^{(k)}$, $\CG^{(l)}$ and $\CG^{(k+l)}$
respectively.

For $\CG=A_1$, there are general perturbative results for large $l$
matching those already
described for the minimal models\ts\RF\rCSSa\CSSa. There is a surprise
here, in that the perturbation of the $l\th$ model of the $k\th$ series does
not flow to the $l{-}1\th$ model of this series, but rather to the
$l{-}k\th$. As a result, there is no longer a single sequence of one-hop
trajectories, but rather $k$ disconnected sequences, interlaced along
the $k\th$ series. Note that taking $\CG=A_1$, $k=2$ gives the
$N{=}1$ superconformal discrete series, for which this result had
already been found in ref.\ts\RF\rPg{\Pg\semi\KIKKMOa}. A TBA analysis for
the general $A_1$ case, for arbitrary values of $l$, has now been
given by \AlZ\ts\RF\rZn\Zn; the picture outlined above holds good modulo a
small complication when $l$ becomes smaller than $k$, which will be
described later. The further generalisation to
arbitrary $\CG$ can be found in ref.\ts\rRb.

Since for given $k$ there are $k$ different sequences of hopping flows,
it is reasonable to hope for $k$ different staircase models
approximating them. Some nice recent work by
Martins\ts\NRF\rMo\Mo\NRF\rMp\Mp\refs{\rMo,\rMp}\ has
started this programme by proposing a generalisation of Zamolodchikov's
original staircase model to one of the $k=2$ sequences,
finding TBA systems indicative of RG trajectories which pass close by
the subset of superconformal minimal models with non-zero Witten index
($l$ even), before flowing off to  massive theories in the IR limit.
He also showed the generalisation of this to
other $\CG$, again at $k=2$ and $l$ even. One interesting
feature of his proposal is that each TBA system contains magnonic
terms (pseudoenergies with no direct couplings to energy terms),
perhaps
indicating that the as-yet unknown scattering theories underlying
the models have non-diagonal S-matrices.

In this paper, we propose and start to analyse
TBA systems for general $k$ and $\CG$. The proposal itself is given in the
next section: for each $k$ there are $k$
different systems, labelled by an index $\w\in\,$\ZK.
In sections 3 and 4, we
describe why these systems should mimic the desired hopping
behaviour, a discussion that has many parallels to one given
earlier for the $k=1$ series\ts\rDRa. The conclusions of these two
sections have been backed up by some numerical work, which is reported
in section 5. The RG flows predicted have a number of surprising
features, which can however be supported by alternative arguments.
These points, along with some further speculations, are contained in
the concluding section.

\newsec{The spiral staircase}

Take $\CG$ to be an arbitrary simply-laced Lie algebra, with Coxeter
number $h$, and
let $a$ label nodes on the corresponding Dynkin diagram. In
real-coupling affine Toda field theory, the different types of
particle are labelled by just such an index; let $m_a$ be the mass of
the corresponding particle. These masses can also be characterised as
the components of the Perron-Frobenius eigenvector of the $\CG$ Cartan
matrix\ts\NRF\rBCDSc\BCDSc\NRF\rKMa\KMa\NRF\rFb\Fb\refs{\rBCDSc{--}\rFb}.
To provide for some magnonic structure, an extra label $i$ will be
needed, lying in \ZK. As already mentioned, the particular staircase out of
the $k$ possibilities is determined by a member of \ZK, $\w$ say, which
will stay fixed throughout the discussion.

The TBA system will be given in terms of $r\times k$
pseudoenergies $\ep^{(i)}_a$ ($r$ is the rank of $\CG$). The
\ZK -condition on the index $i$ amounts to identifying $\ep^{(i)}_a$ with
$\ep^{(i+k)}_a$. Similarly there are $r\times k$ energy terms $\nu^{(i)}_a$,
given by the formula
\eqn\enterm{ \nu^{(i)}_a(\t)=
 \hat m_a(\delta^{i,0}e^{x-\t}+\delta^{i,\w}e^{x+\t})\, ,}
where $\t$ is the rapidity, and $x$ encodes the scale of the system via
$e^{x}={1\over 2}m_1R$. The case $\w{=}0$ will turn out to have a
massive infrared limit, allowing $m_1$ to be interpreted as
the mass of a particle, or multiplet of particles;
the values of the remaining $m_a$ are contained in the
(dimensionless) ratios $\hat m_a=m_a/m_1$. Note that the energy term
is only nonzero if $i$ is equal to $0$ or $\w$: it is in this way that the
value of $\w$ enters the game.  Defining
$L^{(i)}_a(\t)=\log(1+e^{-\ep^{(i)}_a(\t)})$, and introducing a
parameter $\t_0\in\hbox{\R}^+$, the proposed
TBA system is:
\eqn\tba{\ep^{(i)}_a(\t)+{1\over 2\pi}\sum_{b=1}^r\Bigl[
\phi_{ab}*L_b^{(i)} - \psi_{ab}*L^{(i-1)}_b(\t-\t_0)
-\psi_{ab}*L^{(i+1)}_b(\t+\t_0)\Bigr]=\nu^{(i)}_a(\t)\, ,}
where all the $i$-type indices are to be taken modulo $k$, and $*$ denotes
the rapidity convolution: $\phi*L(\t)=\intt d\t'\phi(\t-\t')L(\t')$.
The kernel functions $\phi_{ab}$ and $\psi_{ab}$ are defined in terms
of the minimal parts of the corresponding affine Toda S-matrix
elements $S_{ab}^{\rm min}$, written in the form
$$S^{\rm min}_{ab}(\t)=\prod_{x\in A_{ab}}\(x-1\)(\t)\(x+1\)(\t)\quad ;
\quad \(x\)(\t)={\sinh({\t\over 2}+{i\pi x\over 2h})\over
                 \sinh({\t\over 2}-{i\pi x\over 2h})}\, ,$$
where $A_{ab}$ is a set of integers (possibly with repetitions), and the
related functions
$$S^F_{ab}(\t)=\prod_{x\in A_{ab}}\(x\)(\t)\, .$$
Then
$$\phi_{ab}(\t)=-i{d\over d\t}\log S^{\rm min}(\t)\quad ;\quad
  \psi_{ab}(\t)=-i{d\over d\t}\log S^F(\t)\, .$$
More explanation can be found in refs.\ts\refs{\rRb,\rDRa}; while
this notation is consistent with that of these two references, it
differs from that of ref.\ts\rMo. A complete list of the functions
$S^{\rm min}_{ab}$ can be found in ref.\ts\rBCDSc, and general formulae
in ref.\thinspace\RF\rDf\Df.

The ground-state scaling function is now given in the standard way\ts\rZg\
in terms of the solutions to the TBA system \tba,
and the energy terms \enterm:
\eqn\cfn{c(x,\t_0)={6\over \pi^2}\sum_{a=1}^r\sum_{i=0}^{k-1}
\int_{-\infty}^{\infty}d\t\hat m_a%
e^{x}\nu_a^{(i)}(\t)L^{(i)}_a(\t)\, .}

The next section will discuss the solutions to \tba, explaining how
the scaling function $c(x,\t_0)$ can be expected to behave as a function
of $x$. Before giving these details, it is worth making a few general
comments on the system defined by \enterm\ and \tba. First, notice that
just as in refs.\ts\refs{\rMo,\rMp}, there are many
`magnonic' pseudoenergies with zero energy term: this may be a sign
that for these models too, the underlying S-matrix is non-diagonal.
More striking is the form that the shifted terms in \tba\ take: while
the term shifted by $-\t_0$ couples with $L^{(i-1)}_b$, that shifted
by $+\t_0$ couples with $L^{(i+1)}_b$. Thus these factors are not a
simple continuation of the affine Toda Z-factors, as was the case for
previous staircase TBA systems\ts\refs{\rZo,\rMl,\rDRa,\rMo,\rMp}.
Despite the apparent asymmetry that this implies, the solutions of \tba\
{\it are} symmetric under $\t\rightarrow -\t$, so long as this
transformation is accompanied by an exchange of the pseudoenergies
$\ep^{(i)}_a\rightarrow\ep^{(\w-i)}_a$. The possibility to implement
the parity transformation in this way is a reflection of the
symmetries of the affine $\hat A_k$ Dynkin diagram on which the \ZK\
labels $i$ live, and is reminiscent of the discussion for systems
based on non-affine graphs given in ref.\ts\RF\rRTVa\RTVa. It also
allows the `left-moving' and `right-moving' energy terms in \enterm\
to be coupled to different pseudoenergies for $\w{\neq}0$, a hint
that in these cases the system has a massless
infrared limit. More detailed analysis in a later section will
confirm this expectation.

The most important feature of \tba\ is the non-local nature of the
interactions between the different pseudoenergies. The functions
$\phi(\t)$ and $\psi(\t)$ are exponentially damped away from a
region of order one about $\t{=}0$, so their convolutions $\phi*L(\t)$
and $\psi*L(\t)$ are, up to exponentially small corrections,
dominated by the values of $L(\t')$ in a region of order one about
$\t'{=}\t$. (While functions $L(\t)$ {\it could} be found for which this
does not hold, suffice it to say that these do not seem to
arise amongst the solutions
to \tba.) Hence the three convolution terms in \tba, which involve
$\ep^{(i)}_a$, $\ep^{(i-1)}_b$ and $\ep^{(i+1)}_b$, pick up most of
their values near $\t$, $\t{-}\t_0$, and $\t{+}\t_0$ respectively.
To visualise this, it is helpful consider the index
$i$ on the same footing as $\t$, albeit only taking discrete values.
Thus we write $\ep^{(i)}_a(\t)=\ep_a(i,\t)$, the pair $(i,\t)$ being
valued in \ZK$\times$\R. This space has the form of a cylinder, on
which \tba\ induces interactions between the neighbourhoods of
$(i,\t)$, $(i{-}1,\t{-}\t_0)$ and $(i{+}1,\t{+}\t_0)$. This explains
the sobriquet `spiral staircase': the pseudoenergies
couple together in a spiral pattern round the cylinder, a feature
that turns out to be important in reproducing the expected
hops in the RG flow.

\newsec{The double helix}
We now turn to the solutions to \tba, and to progress we will have
to make some more assumptions about
their general form. These parallel the assumptions made in
the analysis of more usual TBA systems, and we will not pretend to
give any rigorous proofs. However,
we have also made a number of numerical checks, which will be
commented on later.

Comparing \tba\ and \enterm, there are two regions where
$\ep_a(i,\t)$ and $L_a(i,\t)$ are immediately known (as
always, up to exponentially small corrections -- we will tend to
assume such a phrase to apply globally from now on). For $i=0$ and
$\t\ll x$, the energy term $\nu_a(0,\t)$ becomes very large and
dominates \tba, so that $\ep_a(0,\t{\ll}x)\approx \hat m_ae^{x-\t}$.
Correspondingly, $L_a(0,\t{\ll}x)$ suffers a double-exponential decay
and is soon negligably small. In the region $i=\w$, $\t\gg -x$, similar
considerations show that $\ep_a(\w,\t{\gg}-x)$ grows
exponentially, while $L_a(\w,\t{\gg}-x)$ quickly decays towards zero.

Near to $(0,x)$, all the terms in \tba\ come into near-equal
competition, and it is no longer a good approximation to ignore the
convolutions. However, beyond this transitional region the form of the
equation simplifies once again, as for $\t\gg x$ the energy term
$\nu_a(0,\t)$ becomes exponentially small and can be dropped.
On the basis of previously-studied TBA systems, we expect $L_a(0,\t)$
to have a kink near $\t=x$, interpolating between zero for $\t\ll x$,
and some other constant for $\t\gg x$. Just as is the case for
other TBA systems, the precise form
of this kink is hard to find due to the increased complexity of
\tba\ in the transitional region, but is not required for an asymptotic
evaluation of the ground-state scaling function.

At this point the non-locality of \tba\ comes into play.
The presence of a kink near $(0,x)$ has an
effect, via the shifted convolution terms, on the equations obeyed
near $(k{-}1,x{-}\t_0)$ and $(1,x{+}\t_0)$.
This is similar to the propagation of kinks in the $k{=}1$
staircase model described in \rDRa, though with one important
difference: here, the influence is on pseudoenergies with different values
of $i$ from that of the \lq seed' kink near $(0,x)$, and hence with
{\it different} energy terms derived from \enterm.
In particular, while $L_a(0,\t)$ was forced to be zero for
$\t\ll x$ by the dominance of $\nu_a(0,\t)$ in
this region, this is generally not the case for $L_a(k{-}1,\t)$.
Thus we expect to find kinks generated in {\it both} directions from
the initial kink. These in turn
cause there to be kinks at $(k{-}2,x{-}2\t_0)$ and $(2,x{+}2\t_0)$, and
so on, the set of secondary kinks spiralling round the cylinder in
both directions from $(0,x)$.
Making the transformation $i\rightarrow \w{-}i$, $\t\rightarrow-\t\,$
reveals another set of kinks spiraling away from $(\w,-x)\,$,
and interleaving with the first spiral to form the pattern of a double
helix. This is illustrated in figure~1, where the kinks should be
imagined to be strung out like beads along the two spirals. The
two regions where the energy terms are dominant are depicted as double
lines; the two seed kinks are at their ends.
That they do not overlap means that the figure
shows a situation where $x$ is negative.

The asymptotic directions for which $c(x,\t_0)$ can be
evaluated correspond to the kinks from one spiral becoming far away
from those of the other, so that the values of $\ep_a$ and $L_a$ have
time to settle down to approximately constant values in the inter-kink
regions (that they {\it do} become nearly constant in these regions is
in fact the key assumption of the discussion). This breaks
down whenever $(m,x{+}n\t_0)\approx (\w,-x)$ for some $m$.
The (strict!) equality implied on the \ZK-valued first coordinate
requires $m=nk{+}\w$ for some integer $n$, and so interleaving will fail
each time $x/\t_0\approx -(nk{+}\w)/2\,$, corresponding to a crossover
in the critical behaviour. Attention will now be restricted to
situations far from these points, and the scaling function $c(x,\t)$
treated in the limit $\t_0\rightarrow\infty$ with $x/\t_0$ remaining
fixed at a value away from crossover.

While secondary kinks are generated in both directions from each seed
kink, they cannot propagate indefinitely. The fact that
\ZK\ is cyclic means that at some stage energy terms will be
re-encountered in domains where they are dominant, and this
truncates the chain of kinks. For the remainder of this section, we
assume that $x<0$, as in
figure 1; it will turn out that most (though not always all) of the
hops are found in this region. For definiteness, take $x$ to lie between
crossovers in the region defined by
\eqn\ncrossover{
-\bigl(nk+\w\bigr)\t_0 \ll 2x\ll -\bigl((n{-}1)k+\w\bigr)\t_0\, .}
Consider the spiral generated by the kink at $(0,x)$ (the
behaviour of the other spiral will then follow by parity). After
$k{-}1$ steps in the $-\t$-direction, there is a kink near
$(1,x{-}(k{-}1)\t_0)$. The energy term $\nu_a$ is zero in this region,
and so there is no reason for this kink to be suppressed. However,
one more step arrives at $(0,x{-}k\t_0)$. This is well inside the
region $i=0$, $\t\ll x$ where, as already described, the part
$\hat m_a\delta^{i,0}e^{x-\t}$ of the energy term dominates the TBA equation
and forces $L_a(0,\t)$ to be vanishingly small.
There is no kink here -- $L_a(0,\t)$ has been zero since the seed
kink at $(0,x)$, and remains so as $\t$ decreases through $x{-}k\t_0$.
Thus in the $-\t$-direction, the chain of kinks terminates after
$k{-}1$ steps. In the $+\t$ direction, the truncation is instead
effected by $\hat m_a\delta^{i,\w}e^{x+\t}$, the
part of the energy term which couples when $i=\w$. If $\t\ll -x$ this
term plays no r\^ole in \tba, and for $x$ in the region \ncrossover\
the spiral makes $n$ complete circuits of the cylinder, and then
continues with a further $\w{-}1$ kinks, before the energy term
becomes important. So, there are $nk{+}\w{-}1$ kinks in the
$+\t$-direction from the seed kink at
$(0,x)$, truncation being caused by the suppression of the putative
kink at $(\w,x+(nk{+}\w)\t_0)$.

The full picture for both spirals can
be seen by referring once again to figure 1: the spirals terminate
when they encounter the double lines representing regions in which the
energy term is completely dominant. Since they each make a single turn in
one direction, and two complete turns in the other,
the figure should correspond to taking $n{=}2$ in \ncrossover: this
is easily verified by writing down the inequalities implied by the
ordering of the marked points along the $\t$-axis.

We can now unwind each spiral from the cylinder, and represent the set
of interacting kinks graphically:
\medskip
\eqn\diagone{\hbox{\massless{1}\dlink%
\massless{k-1}\link\massive{k}\link\massless{k+1}\dlink%
\massless{}\link\massless{k+nk+\w-1}}}
\smallskip\noindent
Each node is a kink, and nodes are linked if they
interact via the shifted convolution terms in \tba. Empty nodes
($~\bigcirc~$) represent kink regions where the energy term plays no
part in the leading behaviour of \tba, while the filled node
($~\bigotimes~$) corresponds to the seed kink, in the neighbourhood of
which the energy term $\hat m_a\delta^{i,0}e^{x-\t}$ cannot be
ignored. This is also the only region of those represented on the
graph which contributes directly to the formula \cfn\ for
$c(x,\t_0)$. An isomorphic (reflected) graph results for the other
spiral.

The next task is to find the values
taken by the functions $\ep_a(i,\t)$ and $L_a(i,\t)$ in the
inter-kink regions. The necessary constraints follow in the
usual way by pulling near-constant terms out of the
convolutions in \tba, leaving only overall integrals of $\phi_{ab}$
and $\psi_{ab}$. These are\ts\refs{\rKMa,\rRb}:
\eqn\Nvals{ {1\over 2\pi}
\int^{\infty}_{-\infty}d\t\p_{ab}(\t)=\delta_{ab}-2C^{-1}_{ab}
\quad ;\quad {1\over 2\pi}
\int^{\infty}_{-\infty}d\t\psi_{ab}(\t)=-C^{-1}_{ab},}
where $C_{ab}$ is the (non-affine) $\CG$ Cartan matrix.
The various inter-kink regions influence each other through the non-local
terms in \tba, in much the same way as did the kinks.
To describe the situation, let the $p\th$ kink in \diagone\ be
located near $(i_p,\t_p)=(p\,\mod\,k,x{+}(p{-}k)\t_0)$,
and let $(i_p,\t_p^-)$ and $(i_p,\t_p^+)$ be points in the inter-kink
regions immediately before, respectively after, this kink:
\eqn\JohnSelwynGummer{\t_p^-={1\over 2}\t_0\(2p-(n{+}2)k-s\)\quad;\quad
                      \t_p^+={1\over 2}\t_0\(2p-(n{+}1)k-s\)\, .}
Via \tba, the $\ep_a(i_p,\t_p^-)$ come into interaction, as do the
$\ep_a(i_p,\t_p^+)$. However care is needed when $p{=}k$.
While the seed kink at $(i_k,\t_k)=(0,x)$ owes its very
existence to the delicate balance between the energy and convolution
terms in \tba\ at that point, the energy term has become negligable by
the time that $(i_k,\t_k^+)$ is reached, and so has no effect on the
equation for $\ep_a(i_k,\t_k^+)$.  On the other hand,
$(i_k,\t_k^-)$ is in the region immediately to the left of the
seed kink, where the energy term in \tba\ is dominant. As a
consequence $\ep_a(i_k,\t_k^-)$ is forced
 to be effectively infinite, irrespective of
the values taken by the other $\ep_a(i_p,\t^-_p)$.

All this has the effect of changing the `connectivity structure' that for the
kink interactions was summarised in \diagone. It can be visualised on
figure 1 by mentally shifting the spiral based at $(0,x)$ slightly to
the left, and then slightly to the right. Shifting leftwards, the
spiral is cut by the double line representing a dominant energy term,
while shifting rightwards disconnects the spiral
from this line altogether, apart from the endpoint. Thus for
the inter-kink regions containing the left-shifted points
$(i_p,\t_p^-)$, \diagone\
should be replaced by
\medskip
\eqn\diagtwo{\hbox{\massless{1}\dlink%
\massless{k-1}\phantom{\link\massive{k}\link}\massless{k+1}\dlink%
\massless{}\link\massless{k+nk+\w-1}}}
\smallskip\noindent
while to the right, the graph for the $(i_p,\t^+_p)$ is
\medskip
\eqn\diagthree{\hbox{\massless{1}\dlink%
\massless{k-1}\link\massless{k}\link\massless{k+1}\dlink%
\massless{}\link\massless{k+nk+\w-1}}}
\smallskip
For the second spiral making up the double helix, kinks are found at
$(\tilde\imath_p,\tilde\t_p) = (\w-i_p,-\t_p)$ and the two neighbouring
inter-kink regions are centred on $(\tilde\imath_p,\tilde\t_p^{\pm})$,
where $\tilde\t^+_p=-\t^-_p=\t^-_{(n+2)k+s-p}$ and
$\tilde\t^-_p=-\t^+_p=\t^-_{(n+1)k+s-p}$. The above treatment goes through
isomorphically when phrased in terms of these kinks, as indeed it must by
the parity symmetry of the system.

To express the algebraic content of \diagtwo\ and \diagthree, let
$l_{pq}^-$ and $l_{pq}^+$ be their respective incidence matrices,
and put $\tilde C_{pq}^{\pm}=2\delta_{pq}-l_{pq}^{\pm}$. At
$(i_k,\t^-_k)$, the value of $\ep_a$ has already been established; for the
rest, extracting the near-constant values of
$\ep_a(i_p,\t_p^{\pm})$ from the convolutions in \tba\ and using \Nvals\
yields, after some simple rearrangements, the following consistency
conditions:
\eqn\JohnMajor{f_{ap}^{\pm}
 =\prod_{b=1}^r\prod_{q=1}^{\scriptscriptstyle (n+1)k+\w-1}%
\(1-f_{bq}^{\pm}\)^{C_{ab}^{-1}\tilde C_{pq}}\,,}
where
\eqn\MrsThatcher{f_{ap}^{\pm}=
{e^{-\ep_a(i_p,\t_p^{\pm})}\over 1+e^{-\ep_a(i_p,\t_p^{\pm})}}\, .}

These equations reproduce the constraints for the limiting values of
the pseudoenergies in the \ccoset k {nk+\w} {(n+1)k+\w}\
TBA system given in \rRb,
providing the substitution $\exp(-\ep_a(i_p,\t_p^{\pm}))=Y^a_p(\mp\infty)$
is made. While this is suggestive, caution is needed, since
\JohnMajor\ describes the values of the functions $\ep_a$ at various
points along the $\t$-axis, rather than just at $\pm\infty$ as for more
usual systems. Nevertheless, asymptotically far from the crossovers the
different kinks should become decoupled and behave like independent
functions, their only interactions being those explicitly marked
on \diagone. These reproduce exactly the interactions between different
pseudoenergies found in the TBA systems of \refs{\rZn,\rRb}, so we
certainly
expect the evaluation of the function $c(x,\t_0)$ in the asymptotic
regime under discussion (given by \ncrossover) to be the same as for
the corresponding coset TBA system. But to confirm this, the leading
asymptotic behaviour of $c(x,\t_0)$ away from crossover can also
be calculated directly
from the formula \cfn, remaining within the spiral staircase model.
The necessary modifications to the usual arguments have already been
described at some length in ref.\ts\rDRa, and the slightly
increased complexity of the models being considered here has no
bearing on the calculations once the spirals have been unwound, and the
connectivity structure \diagone\ established. The one small point to
watch is that since the massive kink is now located at the $k\th$
position, inside the chain,
the `integrations by parts' by means of which the other kinks are felt
now proceed in {\it both} directions away from $k$, rather than just
to the right as for the $k{=}1$ staircase. Referring to the
earlier paper \rDRa\ for further details, we will simply report
the final result:
\eqn\CecilParkinson{\lim_{{\t_0\to\infty\atop x/\t_0=N_n}}c(x,\t_0)=
c_n={6\over\pi^2}\sum_{a=1}^r\sum_{q=1}^{\scriptscriptstyle (n+1)k+\w-1}%
\[\rdilog\(f_{ap}^+\)-\rdilog\(f_{ap}^-\)\]\,,}
where $x/\t_0$ is held fixed at $N_n$,
$-(nk{+}\w)/2 \ll N_n\ll -((n{-}1)k{+}\w)/2\,$,
while the limit is taken in order to stay inside the regime defined
by \ncrossover. The functions appearing on the righthand side are
Rogers dilogarithms, defined by $\rdilog(z)=-{1\over 2}%
\int_0^zdt\({\log(1-t)\over t}+{\log t\over 1-t}\)\,$, and
their arguments $f_{ap}^{\pm}$ are obtained by
solving \JohnMajor. To finish the calculation, a sum rule is
needed\ts\RF\rBRa{\Kg\semi\BRa}:
if $\{f_{ap}\}$ ($1{\leq}a{\leq}r$, $1{\leq}p{\leq}l$) is the solution
to \JohnMajor\ when $C_{ab}$ is the Cartan matrix of $\CG$ and
$\tilde C_{pq}$ the Cartan matrix of the algebra $A_{r'}$, then
\eqn\SaraKeays{ {6\over\pi^2}\sum_{a=1}^r\sum_{p=1}^{r'}
\rdilog\(f_{ap}\)={rr'h'\over h+h'}=rh'-{r(h{+}1)h'\over h+h'}\,,}
where $h$ is the Coxeter number of $\CG$ and $h'{=}r'{+}1$ the Coxeter
number of $A_{r'}$. The second version is useful because it
shows the sum to be $rh'$ minus the central charge of the $\CG^{(h')}$
WZW model. For the case in hand, the sum over the
$\rdilog(f^+_{ap})$ follows from \SaraKeays\ with $h'=(n{+}1)k{+}\w$,
since \diagthree\ is the Dynkin diagram of $A_{(n+1)k+\w-1}\,$.
For the
$\{f^-_{ap}\}$ there are two disconnected parts to \diagtwo, the
Dynkin diagrams of $A_{k-1}$ and $A_{nk+\w-1}$. Correspondingly the
dilogarithm sum splits into two, summed by taking $h'$ equal to $k$
and $nk{+}\w$ respectively. The absent $k\th$ node does not contribute
since $\rdilog(f^-_{ak})=\rdilog(0)=0$ for all $a$, and so
\eqn\endresult{
c_n=r(h{+}1)\[{k\over h{+}k}+{nk{+}\w\over h{+}nk{+}\w}
-{(n{+}1)k{+}\w\over h{+}(n{+}1)k{+}\w}\]\,,}
the central charge of the \ccoset k {nk+s} {(n+1)k+s}\ coset model.

\newsec{The infrared limit}
To complete the picture of the asymptotics of the ground-state scaling
function $c(x,\t_0)$, the condition that $x$ be negative,
imposed thoughout the last section,
must be relaxed. Since $x$ is related to the physical
parameters of the model by ${1\over 2}m_1R=e^x$, this
corresponds to examining the system in the infrared.

The point to note is that for $x$ positive the nature of the
relationship between the spirals of kinks and the two regions of
energy-term dominance, illustrated in figure 1 for $x$ negative, may
change.  When $x$ becomes positive, the two regions, marked by double lines
on figure 1, start to overlap, as in one region $\t$ runs from $-\infty$
to $x$, in the other from $-x$ to $+\infty$. If $\w=0$, then the
two regions actually collide and this has the effect of killing off the
two seed kinks
completely. Thus in this case the IR limit is simple:
a theory with $c=0$, with the final crossover to a massive phase
occurring near $x=0$. Otherwise, kinks will continue to be
found near $(0,x)$ and $(\w,-x)$ for arbitrarily large positive
values of $x$. Thus the $L^{(i)}_a(\t)$ never become trivial, and a
non-zero value for $c(x,\t_0)$ is expected even as
$x\rightarrow+\infty$. This reinforces the assertion made in section 2
that for $\w\neq 0$ the IR limit of \tba\ should be massless.
Precisely which conformal field theory this massless limit should be
brings one last surprise. In the previous discussion
it was seen that for $x{<}0$ the spiral
generated at $(0,x)$ truncates after a single turn in the $-\t$
direction, when it re-encounters the $i{=}0$ energy term.
This is why
the left-most piece of \diagtwo\ has $k{-}1$ nodes, and one of the
elements of the coset whose central charge \CecilParkinson\ reproduces
has level $k$. But if $x$ is positive, this can change. In the
$-\t$ direction, the chain from $(0,x)$ might encounter the energy
term at $i{=}\w$ {\it before} returning to $i{=}0$.
The onset of this phenomenon is signalled by a crossover
around $x={1\over 2}(k{-}\w)\t_0$, beyond which point
the chain of kinks anchored at $(x,0)$ is terminated in {\it both}
directions by the $i{=}\w$ energy term. It is
straightforwardly seen that this is the last such crossover expected
on the basis of changes to the overall kink structure, and that
after this point the
situation stabilises, the form of the pseudoenergies
remaining unchanged apart from simple translations as
$x\rightarrow+\infty$. The final kink system, governing the
IR limit, can be unwound and represented graphically just as
before. The picture \diagone\ becomes:
\medskip
\eqn\diagone{\hbox{\massless{\w{+}1}\sdlink%
\massless{k-1}\link\massive{k}\link\massless{k+1}%
\ssdlink\massless{k{+}\w{-}1}}}
\smallskip\noindent
where to ease comparison with the earlier graphs, the $\mod$-$k$ values
of the labels have been preserved. This graph has $k{-}\w{-}1$ massless
nodes to the left, $\w{-}1$ to the right. The remaining calculations
now go through unchanged, \CecilParkinson\ becoming
\eqn\blob{\lim_{{\t_0\to\infty\atop x/\t_0=N_{-1}\gg(k-s)/2}}c(x,\t_0)=
c_{-1}= r(h{+}1)\[{k{-}\w\over h{+}k{-}s}+{\w\over h{+}\w}
-{k\over h{+}k}\]\,,}
which is the central charge of the \ccoset {k-s} s k\ coset model.

In this last equation, the fact that $N_{-1}$ is positive means that $x$
tends to $+\infty$ as the limit $\t_0\rightarrow\infty$
is taken, but this does not spoil the
validity of the asymptotic estimates being made. However, the real
interest all along has been to trace the variation in $c(x,\t_0)$ in one
particular model, for which $\t_0$ is fixed. Thus a change in the
point of view is needed to apply the results found so far. If the value of
$\t_0$ is large enough, then as $x$ varies, the pair $(x,\t_0)$ will
pass through a series of regions for which the asymptotic results
\CecilParkinson\ and \blob\ are good approximations to the true values
of $c(x,\t_0)$.  Hence this function will run through the
values $c_n$ in turn, deviating significantly from these numbers
only in the crossover regions. An order-of-magnitude estimate for the
size of these regions is easily given: the approximations leading up
to \CecilParkinson\ and \blob\ were good,
up to exponentially small corrections,
so long as the various kinks in each pseudoenergy $\ep_a^{(i)}(\t)$
were clearly separated along the $\t$-axis. These kinks
have a size of order one (the precise value will depend on the model,
but in any case we are only interested in orders of magnitude in
comparison with $x$ and $\t_0$ here), and so the crossover will start
when the expected positions of two kinks become closer than this. There
are generally two different interkink separations for any
given $(x,\t_0)$ (this is rather clear from looking at figure 1); they are:
\eqn\Gould{
\eqalign{&\t_p-\tilde\t_{(n+1)k+\w-p}=-2x-((n{-}1)k+\w)\t_0\,;\cr
 &\tilde\t_{(n+1)k+\w-p}-\t_{p+k}=2x+(nk+\w)\t_0\,.\cr}}
(That these two are positive follows from \ncrossover; strictly
speaking there are also steadily growing separations for $x$ larger
than the final crossover, after the two spirals have become completely
disentangled.) The interkink separations are therefore overly small only
in regions of order one about each crossover value of $x$. Since the
intervals between these crossover values grow linearly with $\t_0$, the
staircase-like nature of $c(x,\t_0)$ soon becomes pronounced. This is
the evidence for the previously-advertised roaming RG trajectories,
and from the values of $c_n$, the set of fixed points approached by
any particular flow can be read off. The \coset k l\ coset models can
be imagined to be located on a grid, giving the following skeleton for
the large-$\t_0$ pattern of the $k,\w$ ($\w{\neq}0$) flow:
\eqn\ksflow{\eqalign{
            & \quad\downarrow\cr
            & c(k,2k+\w)\cr
            & \quad\downarrow\cr
            & c(k,k+\w)\cr
            & \quad\downarrow\cr
  c(k-\w,\w) \leftarrow\ &c(k,\w)\cr}}
The set-up for $\w{=}0$ is less unexpected:
\eqn\kszflow{\eqalign{
            & \quad\downarrow\cr
            & c(k,2k)\cr
            & \quad\downarrow\cr
            & c(k,k)\cr
            & \quad\downarrow\cr
            & \quad 0 \hbox{ (massive)}\cr}}
Before attempting to interpret these results, it is worth seeing that
they stand up to numerical verification.

\newsec{Numerical work}
The above has relied rather heavily on assumptions about the behaviour
of the solutions to \tba\ -- the existence of kinks and so on --
which, though well-motivated, have most certainly not been rigourously
derived. Therefore it is worth subjecting the proposal to some
independent checks, and for this we have solved the equations \tba\
numerically in a number of cases, discretising the $\t$ axis at
intervals of $0.2$, and then iterating \tba\ until $c(x,\t_0)$
relaxed to a steady value. To gain five-digit precision, ample for the
purposes of graph-plotting, typically took from 25 to 30 steps.
We have only looked at the case $\CG=A_1$ -- for higher-rank algebras
the iteration of TBA equations, even in the usual cases, is more
tricky\ts\RF\rKMb\KMb\ -- but previous numerical work on higher-rank
staircase models\ts\refs{\rMl,\rMo,\rMp}\ gives no reason to expect
any unpleasant surprises. In particular, in refs.\ts\refs{\rMo,\rMp}\
Martins proposed and investigated numerically
the $k=2$, $\w=0$ instance of \tba\ with $\CG=A_2$, and the
predictions made above for this case are consistent with his findings.

Figures 2a and 2b show numerical results for all values of $\w$ at
$k=2$ and $k=3$, respectively. In both cases $\t_0$ was fixed at $40$,
and it can be seen that the agreement with the predictions of the last
two sections is excellent, even including the final `corner' of
\ksflow\ for $\w\neq0$. One amusing feature is that for $k=3$, the two
flows with non-zero $\w$ have the same infra-red central charge,
despite their very different behaviours at intermediate scales.
At given $k$, each pair of flows $\w$, $k{-}\w$ has this property --
at the simplest level just a reflection of the fact that
c(\ccoset {k-s} s k) = c(\ccoset s {k-s} k). Note though that this is
only an equality of central charges -- since our system in its current
form only traces the evolution of the ground-state energy, we cannot
distinguish between different modular invariants having the same central
charge.

\newsec{Discussion}

The convergence of analytical and numerical results leaves little
doubt that the behaviour of the solutions to \tba\ is as claimed
above,  even if this has not been rigourously proved.
It is then very tempting to suppose that \tba\ is
indeed the TBA system for some relativistic field theory, the
exact ground-state scaling function of which is given by $c(x,\t_0)$.
In the absence of any concrete proposals as to what this model might
be, we can at least discuss some of its expected properties, and
decide whether they are consistent from other points of view.

The first of these properties is that as $x$ increases from the deep
ultraviolet, $c(x,\t_0)$ does indeed run through the
sequence of numbers $c_n$ given by equation \endresult, these being a
subset of the \coset k l\ central charges. The function
pauses near each $c_n$ for a `time'
approximately given by $k\t_0/2$ for large $\t_0$, before making a
sharp transition to $c_{n-1}$ over an interval with
size of order one, in the process of which $l$ decreases by $k$.
This is just the behaviour needed
for a staircase model based on these cosets, being consistent
with the existence of a family of roaming flows which in the
large-$\t_0$ limit comes to approximate the known single-hop trajectories.

More interesting is the predicted behaviour towards the infrared, as
$x$ increases and $l$ becomes smaller. The hopping-by-$k$ cannot
continue indefinitely, since $l$ cannot be negative. For $\w{=}0$, the
situation is simple: after a last pause near $l{=}k$, the final flow
is to $c=0$, implying that the theory becomes massive at long
distances. All previously-studied staircase models have shared this
property, which is furthermore in accord with the one-hop behaviour
predicted for the $\phi_{id,id,adj}$-perturbation of a
\ccoset k k {2k}\ coset model. To understand the curious final hop
that all the staircase systems with $\w{\neq}0$ undergo, we first
recall the general small-$l$ pattern of the
interpolating trajectories\ts\rZn. Even without the TBA, it is
clear that once $l$ has become smaller than $k$,
it is no longer possible for $l$ to decrease by a further
$k$. However the perturbation $\phi_{id,id,adj}$
is symmetrical with respect to the $k$ and $l$ in
the numerator of the coset \coset k l, and so as soon as $l$ decreases
below $k$, it is natural to expect that the flow induced by this
perturbation now hops by
decreasing $k$ by an amount $l$, instead of $l$ by $k$.
Such a change in direction for the
interpolating trajectories when $l$ becomes smaller than $k$ is indeed
predicted by Zamolodchikov's TBA analysis\thinspace\rZn.

For the
superconformal discrete series ($\CG{=}A_1$ and $k{=}2$),
such a phenomenon was discussed in
ref.\ts\RF\rKMSa\KMSa, and can be visualised by referring back to
figure 2a, which shows the form of the two flows predicted by \tba\
for this case. The series of models with $l$ odd continues to hop down
in steps of $2$ until $l{=}1$, the tricritical Ising model, is
reached. All of the models up to and including
this point have been $N{=}1$
supersymmetric, a symmetry respected by their $\phi_{id,id,adj}$
perturbations -- the field is in fact the top
component of a supermultiplet, all other components vanishing anyway
on integration over the Grassman directions. However while the
$\phi_{id,id,adj}$ operator of the tricritical Ising model again
respects the $N{=}1$ supersymmetry, it is also the $\phi_{13}$
operator in the sense of the minimal $c{<}1$ series, and as such
induces a flow down to the Ising model, which does {\it not} possess
such a symmetry. The interpretation given in ref.\ts\rKMSa\ is that
the supersymmetry is spontaneously broken along this last
trajectory.  At large $\t_0$ we can expect the $k{=}2, \w{=}1$
staircase model to show similar behaviour, with an approximate
supersymmetry at short distances being broken near $x{=}\t_0/2$
as the trajectory approximates the final step down to the
Ising model. Thus a change in the hop direction, from decreasing $l$
to decreasing $k$, is not ruled out, and
should probably be associated with the spontaneous breaking of
whatever higher symmetry is associated with a given series. For
$k{=}2$, this was $N{=}1$ supersymmetry, and the situation was
reasonably under control; for large values of $k$, or algebras $\CG$
other than $A_1$, the situation is less clear, and the discussion
much more speculative.
The possible forms of the symmetries associated with such theories were
first discussed in refs.\ts\RF\rRc{\Rc\semi\BNYa\semi\KMQa},
and later in ref.\ts\RF\rBLb\BLb\ where they were called `fractional
supersymmetries',  but in particular the
lack of a satisfactory generalisation of the Grassman variables to
these situations makes life difficult.
A better understanding of all
this will be needed before any detailed implications can be drawn for
the symmetries of the staircase models.

To return to the general form of the staircase flows,
it is tempting to imagine that they should ultimately approximate a
complete sequence of these
single hops, joined up nose-to-tail to form a
zig-zag series of flows with the left and right coset indices
taking turns to decrease, until a diagonal coset ($k{=}l$) is reached,
at which point the flow would be to a massive model. However this
is {\it not} what happens for the spiral staircase flows predicted by
\tba\ and illustrated in \ksflow: after only a single
hop in a new direction, the flow grinds to a halt, the final
destination in the infrared being the \ccoset {k{-}s} s k\ coset. In
terms of the higher symmetries touched on above, the staircase
continues to mimic the sequence of single-hop flows only so long as
the perturbing operator respects the symmetry associated with the
$k\th$ series of cosets -- after the final flow in which the symmetry
is spontaneously broken, there are no longer any directions respecting
this symmetry, and the flow is `trapped'.

For an alternative understanding of why the flow might come to such an
abrupt stop, first recall some work by L\"assig\ts\RF\rLd{\Ld\semi\Le}\
on Zamolodchikov's original staircase model. The hopping flows
mimicked in this case were those between minimal models,
${\cal M}_p\rightarrow {\cal M}_{p-1}$, induced by the massless
$\phi_{13}$ perturbation. L\"assig pointed out that at large $p$ an RG
flow with properties identical to that predicted by Zamolodchikov can
be uncovered within a perturbative treatment, simply by adding
the slightly irrelevant operator $\phi_{31}$ to the original
perturbation by the just-relevant operator $\phi_{13}$.
Normally one ignores irrelevant operators as having no effect on the
infrared destination of an RG trajectory, but this may not be correct if
the situation under discussion involves a crossover in critical
behaviour, as here. The subtlety is that even an irrelevant operator
with respect to the first fixed point may induce couplings to relevant
operators near the second,
repelling the trajectory and sending it on to some new
destination of even lower criticality. A detailed analysis of the
relevant RG equations, to lowest order in $1/p$, shows that this is
indeed what happens for the combined $\phi_{13}$,$\phi_{31}$ flows
in the minimal series -- in fact, both operators are
nearly marginal and mix under the RG flow, so that, crudely speaking,
the irrelevant $\phi_{31}$ operator near ${\cal M}_p$ becomes
$\phi_{13}$ as ${\cal M}_{p-1}$ is approached, and sends the flow
leapfrogging on down the minimal sequence. Thus even
an irrelevant operator can cause an interpolating flow to change its
IR behaviour, a judicious choice managing to replicate itself at the
next step, and ultimately producing a flow with the characteristic
staircase-like form. It is important here that the original
perturbing operator for the interpolating flow, $\phi_{13}$, flowed
precisely to $\phi_{31}$ in the infrared -- had it flowed to anything
else, the subspace of couplings under consideration would have had to
be enlarged at least to include that operator, spoiling the simple
leapfrogging picture. To repeat this analysis for a level-$k$
staircase, it is therefore natural to imagine a \coset k l\ model
perturbed by a linear combination of the relevant operator
$\phi^{(k,l)}_{UV}=\phi_{id,id,adj}$, which on its own
induces the interpolating flow to the next model down, and the
irrelevant operator $\phi^{(k,l+k)}_{IR}$, along which the
trajectory arriving from the model one step higher arrives.
(There is also an incoming flow $\phi^{(k+l,l)}_{IR}$ forming part of
a level-$l$ sequence of hops, but we won't worry about this for now.)
In ref.\ts\rZn, Zamolodchikov found the incoming operator in the case
$\CG=A_1$ to have conformal dimension
\eqn\Cook{\Delta_{IR}=1+{2\over l+2}\,,}
corresponding to the field $\phi_{id,adj,id}$. The
flow into $l{=}1$ is exceptional, in that there is no operator in the
infrared model with this conformal dimension -- the adjoint is not
among the level-one representations of $\hat {SU(2)}$. Instead,
the prediction is that $\Delta_{IR}{=}2$, leading to the expectation
that here the attracting operator is $T\bar T$. Staying with this
case, which for the spiral staircase corresponds to setting $\w{=}1$,
we should examine the effect of adding some $T\bar T$ to the
$\phi_{13}$-perturbation of the $(k,1)$ $A_1$ coset model.
For large $k$ a treatment perturbative in $1/k$ should be
valid. Note that even in this limit the operator $T\bar T$ retains a
scaling dimension of $4$, a feature which already distinguishes this
case from that of the combined $\phi_{13}$, $\phi_{31}$ perturbations.
But more important is the fact that $T\bar T$ is a symmetrical
descendant of the identity, so that the operators generated in
repeated operator product expansions of $\phi_{13}$ and $T\bar T$
only consist of left-right symmetric descendants of $\phi_{13}$. Since
$\phi_{13}$ itself is already nearly marginal in the large-$k$ limit,
any such descendants will be strongly irrelevant. Any operator mixing
induced through the RG flow is restricted to operators in this
subalgebra, and therefore should not affect the final destination of
the flow. Near to the $(k,1)$ fixed point (${\cal M}_{k+2}$),
there are thus (at least) two
different subspaces pertinent to staircases: one spanned by
$\phi_{13}$ and $\phi_{31}$, and one by $\phi_{13}$ and $T\bar T$. The
first of these is relevant to the original ($k{=}1$)
staircase of Zamolodchikov,
as shown by L\"assig, while we expect the second to contain the
tail-end of the $k\th$ $A_1$ spiral staircase model at $\w{=}1$,
undergoing its final crossover before flowing down to the coset model
at $(k{-}1,1)$ in the far infrared. The more detailed analysis made by
L\"assig also gave signs of other flows complementary to the
staircase, obtained by varying the signs of the couplings. It would be
interesting to investigate these possibilities, but for the moment we
content ourselves with this plausibility argument for the form of the
final step of the $\w{=}1$ spiral staircase.

There do not appear to be any obstructions in principle to applying
analogous arguments in the cases $\w{>}1$ and/or $\CG{\neq}A_1$.
Zamolodchikov\ts\rZn\
found that in the case relevant for our $\CG=A_1$, $\w{=}2$
staircases, the incoming direction at the end of the second-last
crossover mimicked by the roaming flow is $G\bar G$,
rather than the $T\bar T$ above, where $G$ is the spin $3/2$ part of
the supercurrent. This is certainly suggestive,
but to give a satisfactory analysis of the general case
requires a more detailed knowledge of the fractional supersymmetry
algebras than we have at present.

A final area for further speculation concerns the field theories
underlying our TBA systems. If such models exist, they must have a
rich and varied structure, and only a subset of them can be massive.
On the basis of the magnonic structure, we would expect to
find non-diagonal scattering in these cases. The $k{=}1$ systems
were intimately related to the real-coupling affine Toda
theories\ts\refs{\rMl,\rDRa}, and so it is possible that further
insight may be found in the fractional supersymmetric sine-Gordon models
introduced by Bernard and LeClair\ts\rBLb.
While we have no definite suggestions to make in this direction, it
seems to be an interesting area for further work.

\bigskip\noindent{\bf Acknowledgements}\smallskip\nobreak
We would like to thank the Isaac Newton Institute for its
hospitality during the completion of this work, and  John
Cardy for useful discussions. PED thanks M\'arcio Martins for
sending a copy of his paper \rMo\ prior to publication.
This work was supported by a grant under the
EC Science Programme (PED), and by the Isaac Newton Institute and INFN
(FR), all of whom we acknowledge and thank.
\medskip

\listrefs
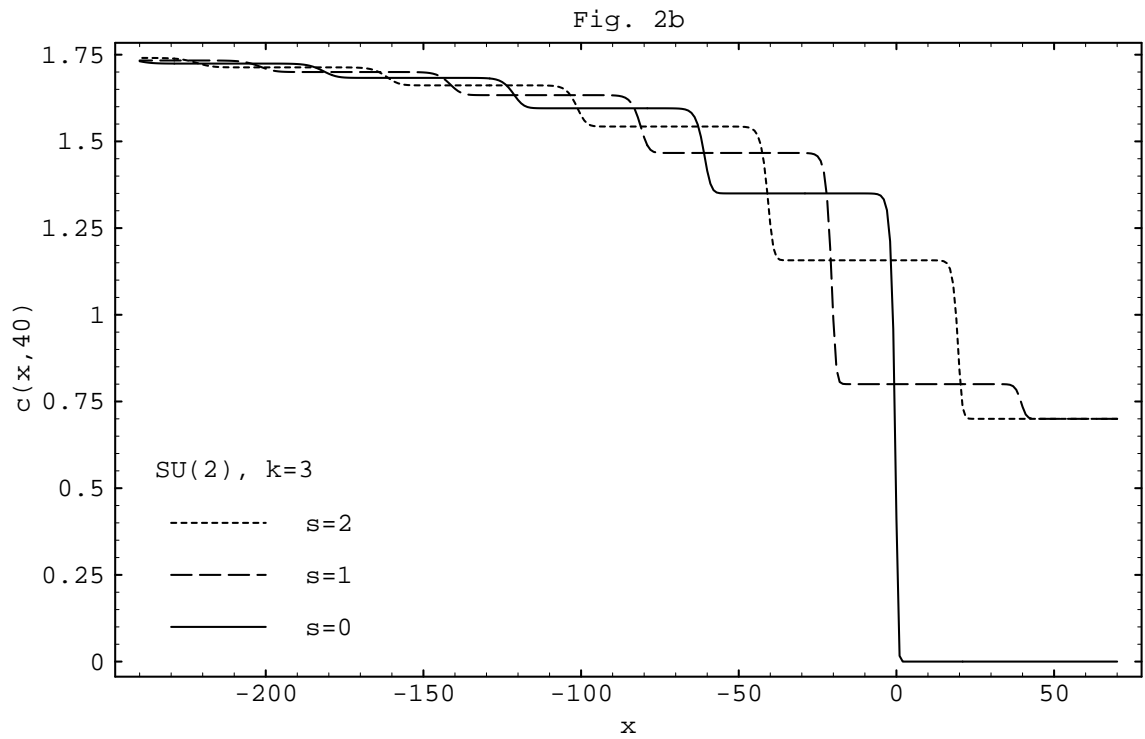
\bye